\documentclass[journal,twoside,a4paper]{IEEEtran}

\usepackage{color}
\usepackage{amssymb}
\usepackage{rotating}
\usepackage[T1]{fontenc}
\usepackage{lscape}
\usepackage{dblfloatfix}
\usepackage{xcolor}
\usepackage{subcaption}
\usepackage[ruled,vlined]{algorithm2e}
\usepackage[autostyle]{csquotes}
\usepackage{tikz}
\usepackage{acro}
\usepackage{siunitx}
\usepackage{enumitem}
\usepackage{multirow}
\usepackage{cite}
\usepackage{hyperref}
\usepackage{amsmath,amssymb,amsfonts}
\usepackage{tikz}
\usepackage{float}
\usepackage{algorithmic}
\usepackage{graphicx}
\usepackage{textcomp}
\usepackage{xcolor}
\usepackage{url}

\usepackage{comment}
\def\BibTeX{{\rm B\kern-.05em{\sc i\kern-.025em b}\kern-.08em
    T\kern-.1667em\lower.7ex\hbox{E}\kern-.125emX}}

\ifCLASSINFOpdf
\else
 
\fi

\usepackage[a4paper, total={170mm,257mm}]{geometry}

\begin{document}

%


\title{An Intelligent Edge-Deployable Indoor Air Quality Monitoring and Activity Recognition Approach}


\author{\IEEEauthorblockN{Mohamed Rafik Aymene Berkani\IEEEauthorrefmark{1},
Ammar Chouchane\IEEEauthorrefmark{2,4}, 
Yassine Himeur\IEEEauthorrefmark{3}, 
Abdelmalik Ouamane\IEEEauthorrefmark{4} and
Abbes Amira\IEEEauthorrefmark{5} 
}\\

\IEEEauthorblockA{\IEEEauthorrefmark{1}
Research Laboratory in Advanced Electronics Systems (LSEA) University Yahia Fares of Medea
Medea, Algeria (berkani.aymene@univ-medea.dz)}\\

\IEEEauthorblockA{\IEEEauthorrefmark{2}University Center of Barika. Amdoukal Road, Barika, 05001, Algeria. (ammar.chouchane@cu-barika.dz)}\\

\IEEEauthorblockA{\IEEEauthorrefmark{3}College of Engineering and Information Technology
University of Dubai, United Arab Emirates (yhimeur@ud.ac.ae)}\\

\IEEEauthorblockA{\IEEEauthorrefmark{4} Laboratory of LI3C, University of Biskra, Algeria (ouamaneabdealmalik@univ-biskra.dz)}\\

\IEEEauthorblockA{\IEEEauthorrefmark{5} Department of Computer Science, University of Sharjah, Sharjah, United Arab Emirates (aamira@sharjah.ac.ae)}\\
}

\maketitle

\begin{abstract}
The surveillance of indoor air quality is paramount for ensuring environmental safety, a task made increasingly viable due to advancements in technology and the application of artificial intelligence and deep learning (DL) tools. This paper introduces an intelligent system dedicated to monitoring air quality and categorizing activities within indoor environments using a DL approach based on 1D Convolutional Neural Networks (1D-CNNs). Our system integrates six diverse sensors to gather measurement parameters, which subsequently train a 1D CNN model for activity recognition. This proposed model boasts a lightweight and edge-deployable design, rendering it ideal for real-time applications. We conducted our experiments utilizing an air quality dataset specifically designed for Activity of Daily Living (ADL) classification. The results illustrate the proposed model's efficacy, achieving a remarkable accuracy of 97.00\%, a minimal loss value of 0.15\%, and a swift prediction time of 41 milliseconds. 

\end{abstract}


\section{Introduction}
Ensuring healthy air quality, particularly in enclosed spaces such as residences and offices, is crucial for safety and well-being. Technological advancements have spurred the development of smart systems that can accurately detect and classify various indoor activities, contributing significantly to environmental safety and quality \cite{himeur2022ai}. These systems employ multiple sensors to measure attributes like temperature, humidity, particulate matter, and volatile organic compounds. The gathered data fuel data-driven models capable of distinguishing between different types of indoor activities \cite{elnour2022performance}. This accurate categorization can trigger appropriate responses, such as activating ventilation systems during cooking to decrease harmful particles or reducing energy consumption during periods of inactivity. Leveraging modern technology and data-driven approaches allows for the creation of intelligent systems that enhance the health, safety, and efficiency of our indoor environments \cite{elnour2022neural,chouchane20183d}.

The field of activity recognition has recently received considerable attention due to its vast applications across various domains such as healthcare, robotics, surveillance, and human-computer interaction ~\cite{khaire2022dl}. Of particular importance is the classification of Activities of Daily Living (ADL), which is instrumental in understanding and predicting human actions. This is crucial for advancements in personalized healthcare, elderly care, and behavioral analysis ~\cite{salguero2019methodology}. The ultimate aim of ADL classification is to enhance an individual's quality of life and foster their independence. Despite being a promising research area, ADL classification presents several challenges such as variability in sensor data, inter-subject and intra-subject variability, and limited sensor coverage, among others ~\cite{elwin2022ar}.

Machine Learning (ML), Deep Learning (DL), and the Internet of Things (IoTs) are powerful tools that can be used to enhance air quality in buildings by classifying ADL. ADLs refer to everyday tasks and activities carried out by individuals within an indoor environment \cite{atalla2023iot,sayed2023edge}. By classifying these activities, we can create more efficient and healthier indoor environments \cite{copiaco2023innovative}.
Firstly, numerous sensors can be installed in a building to monitor air quality and detect various parameters like carbon dioxide levels, temperature, humidity, particulate matter, and volatile organic compounds. These sensors generate large amounts of data, which can be difficult to interpret and use effectively without advanced tools such as ML and DL \cite{sayed2023time}.
ML algorithms, especially supervised learning ones, can be trained on this data, with each activity being labeled according to its impact on indoor air quality. For instance, cooking might be associated with increased levels of certain pollutants. Over time, these algorithms can learn to distinguish between different types of activities based on the changes they cause in the indoor environment.
DL takes this a step further by using neural networks to automatically extract complex features from the raw data \cite{jaleel2023analyzing,belahcene20143d}. A specific type of neural network known as Convolutional Neural Network (CNN) has proven particularly useful for ADL classification \cite{xie2022sequential,katrompas2022recurrence}. CNNs are capable of handling multidimensional data and can learn to identify intricate patterns and temporal dependencies within the data, which might be indicative of specific activities. For instance, a certain combination of changes in temperature, humidity, and pollutant levels might be identified by the CNN as corresponding to cooking activity \cite{salguero2019methodology}.
After identifying the activities, appropriate actions can be taken to mitigate any adverse effects on the air quality. For example, if a DL model identifies an activity that leads to poor air quality, it could trigger a ventilation system to turn on and filter the air. Moreover, the system could also provide real-time feedback to the occupants about their activities and their impact on indoor air quality, thereby encouraging behaviors that maintain good air quality \cite{srivatsan2022ensemble,arrotta2022marble}.

Specifically, DL models have shown substantial effectiveness in the field of air quality monitoring, specifically in classifying and predicting various types of gases and their respective concentrations \cite{jaleel2023analyzing}. By using deep neural networks, researchers can process and analyze large-scale datasets comprising information on indoor gas concentrations over time. Various DL techniques, each with unique advantages and capabilities, have been employed to address the task of air quality monitoring \cite{katrompas2022recurrence,xie2022sequential,chouchane20143d,bessaoudi2021multilinear}.
The development of an intelligent system capable of processing gas detection and monitoring air quality within homes and buildings is of paramount importance. Such a system has the potential to significantly enhance safety, environmental monitoring, and overall well-being within indoor environments. By accurately identifying different types of gases and their concentrations, the system can provide early detection and warnings for hazardous gases. This feature is vital in preventing potential health risks, including respiratory problems, poisoning, and fires, while ensuring that the general air quality meets acceptable standards. Moreover, an intelligent system for gas classification can contribute to energy efficiency by identifying and addressing gas leaks, thereby minimizing waste and potential damage \cite{ertas2022guiding}.

However, the application of DL in this domain is not without challenges due to various reasons. For example, adequate, high-quality data is crucial for training effective DL models. In the context of air quality monitoring, collecting a large amount of accurate and diverse data that represents different gases, their concentrations, and other environmental conditions can be difficult \cite{himeur2021artificial}. Sensor errors, missing data, and data noise are other common problems that can affect data quality \cite{himeur2022recent}. DL models, particularly complex ones like Convolutional Neural Networks (CNNs) or Recurrent Neural Networks (RNNs), require significant computational resources for training \cite{himeur2020novel}.
Additionally, DL models are often considered as "black boxes" due to their complex structure and lack of interpretability \cite{elnour2022neural}.
Moving on, they are sensitive to the data they are trained on. If the training data does not sufficiently represent the diversity of real-world scenarios, the model may not generalize well to new, unseen situations \cite{sayed2022deep}.
Air quality varies over time and across different locations, introducing another level of complexity to the modeling process. Capturing these temporal and spatial patterns effectively in a DL model can be challenging \cite{zhang2022deep}.
The accuracy of predictions also depends on where and how sensors are placed and calibrated. Improper placement or calibration can lead to skewed or inaccurate data, which in turn impacts the performance of the DL model \cite{sardianos2020model}.

To overcome the abovementioned issues, this paper presents an application of CNN-based methods for Activity of Daily Living (ADL) classification, with the primary goal of enhancing the accuracy and efficiency of Air Quality Systems. Leveraging the hierarchical structure of CNNs, which automatically extract complex features, the model can effectively capture distinctive patterns and temporal dependencies from multisensor data.
Typically, the proposed methodology involves developing a novel deep-learning model for classifying various activities, utilizing measurement parameters collected by sensors. To assess the efficacy of our approach, we perform experiments using an air quality dataset specifically curated for ADL classification. The contributions of this paper are twofold. First, we introduce a new DL model that precisely recognizes and classifies four distinct types of air quality based on indoor gas concentration. Second, we evaluate our proposed model, offering insights into its performance and effectiveness.
The principal contributions of this article are summrized as follows:
\begin{itemize}
\item Presenting an intelligent system that is dedicated to monitoring air quality and identifying activities in indoor environments using a DL approach.
\item Introducing a lightweight and edge-deployable 1D-CNN-based activity recognition system, making it suitable for real-time applications and integrating six diverse sensors to gather different measurement parameters.
\item Evaluating the propose method on an air quality dataset specifically developed for ADL classification tasks.
\item Presenting a high level of effectiveness, achieving an impressive accuracy rate of 97.00\%, a minimal loss value of 0.15\%, and a rapid prediction time of 41 milliseconds.
\end{itemize}


\section{Proposed Methodology }
\begin{figure}[t!]
  \centering
   \includegraphics[width=0.40\textwidth]{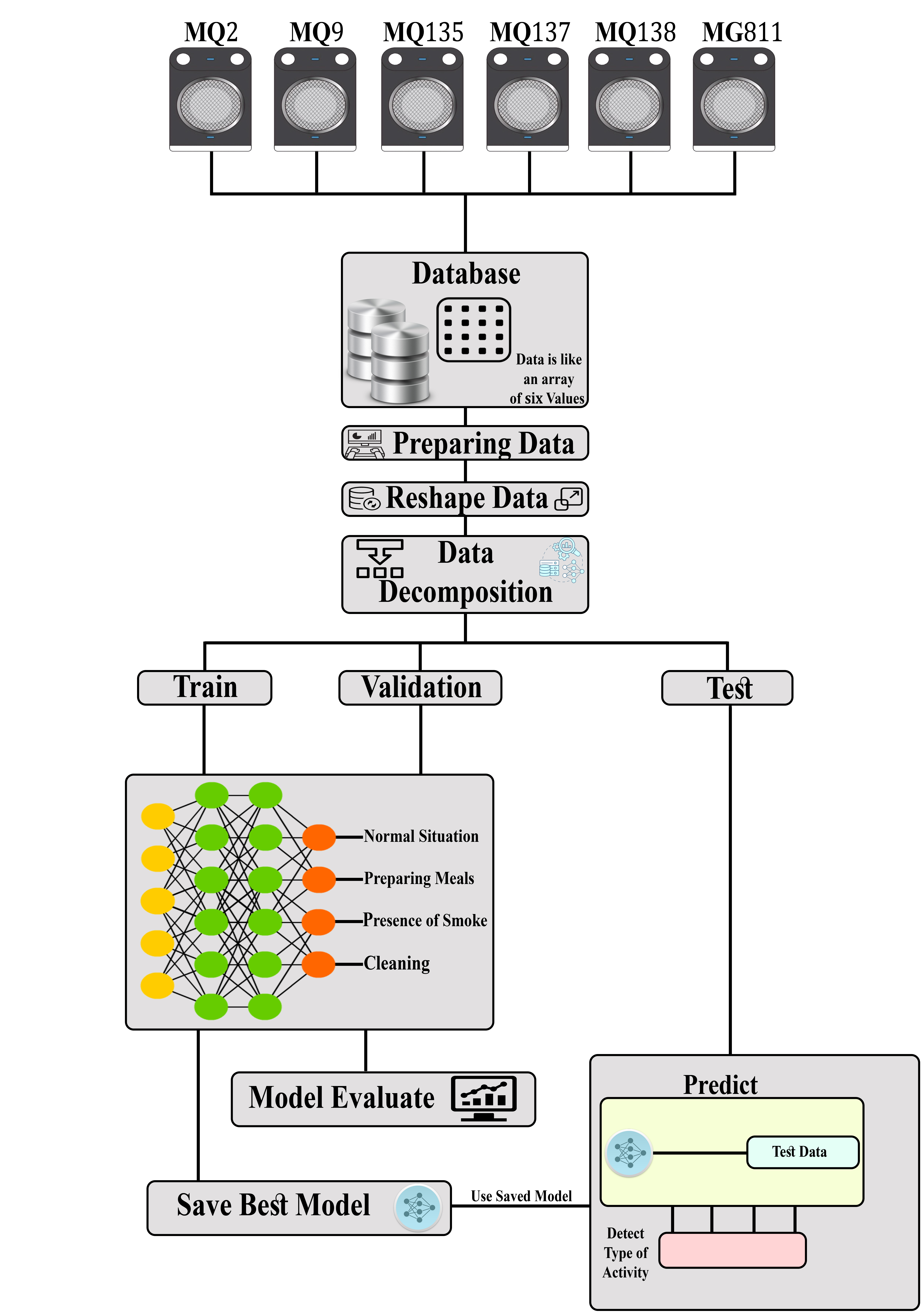}
  \caption{The pipeline of the proposed system}
  \label{fig:fig2}
\end{figure}

This section explains in detail the proposed ADL classification framework. 
Fig. \ref{fig:fig2} presents the proposed ADL classification system which is designed to train and deploy a smart system capable of detecting different activities and air quality by the gas concentration in homes and buildings. 

Initially, there are six different types of sensors (MQ2, MQ9, MQ135, MQ137, MQ138, and MG-811) placed in indoor spaces to collect sensor data. Each sensor provides specific parameters for measuring gas concentration. The collected data is organized into a dataset, typically represented as a table or CSV file. Each line (row) in the dataset corresponds to a specific instance and contains the values from the six sensors, representing their respective measurement parameters.
Each one of the sensors depicted in Fig. \ref{fig:fig2} is designed to detect one or more of the gases specified in Table \ref{tab:table1}. The concentration of gases mentioned in this table can be used as input to our model for the detection of air quality in indoor spaces.

\begin{table}[t!]
\centering
\caption{Sensors and Gas types}
\begin{tabular}{c|c}
\hline
\textbf{Sensors} & \textbf{Gas}                                                                                                       \\ \hline
\textbf{MQ2}     & \begin{tabular}[c]{@{}c@{}}Molecular Hydrogen, LPG, Natural Gas, \\ Carbon Monoxide, Alcohol, Propane\end{tabular} \\ \hline
\textbf{MQ9}     & Natural Gas, LPG, Carbon Monoxide                                                                                  \\ \hline
\textbf{MQ135}   & \begin{tabular}[c]{@{}c@{}}Ammonia, Carbon Mono- and Dioxide, \\ Ethanol, Toluence, Acetone\end{tabular}           \\ \hline
\textbf{MQ137}   & \begin{tabular}[c]{@{}c@{}}Ammonia, Carbon Monoxide, Ethanol, \\ Dimethyl ether\end{tabular}                       \\ \hline
\textbf{MQ138}   & \begin{tabular}[c]{@{}c@{}}n\_Hexane, Benzene, Natural Gas, \\ Carbon Monoxide, Alcohol, Propane\end{tabular}      \\ \hline
\textbf{MG-811}  & Co\_2                                                                                                              \\ \hline
\end{tabular}
\label{tab:table1}
\end{table}

\subsubsection{Reshape Dataset}

To prepare the dataset for training our model, we need to reshape it. Since each line (row) of the dataset contains six sensor values, such as gas concentration, the reshaping process involves converting the dataset into a suitable input format, such as a 2D array. Each reshaped data instance represents a vector with six columns, resulting in a shape of (6, 1).

\subsubsection{Split of the Dataset}
The reshaped data is randomly shuffled and divided into different sets for training, validation, and testing purposes ~\cite{bengherbia2022real}. The split is performed as follows:
\begin{itemize}
    \item Part 1: 70 percent of the data is allocated for the training of our model.
    \item Part 2: The remaining 30 percent of the data
    is further split into two subsets: 20 percent is reserved for validation and fine-tuning of the model during the training process. The remaining 10 percent serves as an independent test set, used to evaluate the final performance of the trained model ~\cite{bengherbia2022real}.
\end{itemize}

\begin{figure}[t!]
  \centering
   \includegraphics[width=0.44\textwidth]{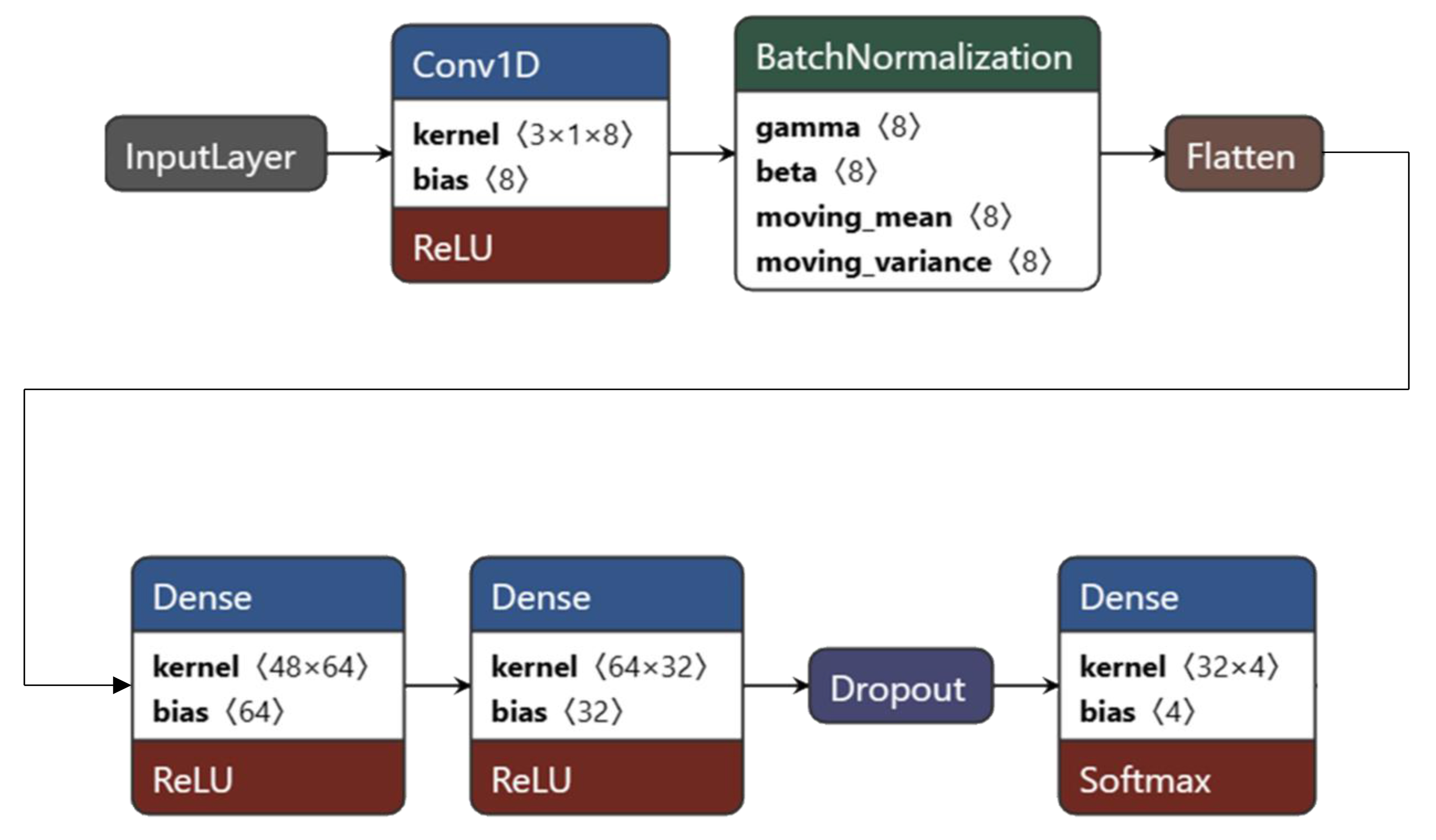}
  \caption{CNN Proposed model.}
  \label{fig:fig9}
\end{figure}

The proposed model architecture, shown in Fig. \ref{fig:fig9}, is a 1D CNN aimed for the classification of diverse in indoor spaces activities and air quality, based on the gas concentration. The model uses a 6-input size, representing the measurement parameters from six sensors (MQ2, MQ9, MQ135, MQ137, MQ138 and MG-811) placed either at home or in a building. The lightweight model is edge deployable to make fast, accurate predictions.

The computational time required to compute a CNN model is typically influenced the number of parameters it possesses. A higher parameter count often corresponds to increase computational complexity. Therefore, the proposed model, it occupies an estimated storage space of 112 Ko only, and contains 5412 parameters.

The proposed model was trained, 
where it learned to recognize the measurement parameter data. Validation set was used during training to assess the model's performance.The validation set is used during the training phase to tune the hyperparameters of the our deep model and monitor its performance. The training process was conducted on Google Colab, employing 200 epochs, a batch size of 64, the adam optimizer, and the categorical-crossentropy loss function.

The deep model was developed and trained using Google Colab, a cloud-based platform for machine and DL applications. The training process involved 200 epochs, which refers to the number of times the entire training dataset was iterated during training. A batch size of 64 was used, indicating that the model processed 64 samples at a time before updating its parameters.Then, Adam optimizer, a popular optimization algorithm, was employed to adjust the model's parameters and optimize its performance. The categorical-crossentropy loss function, specifically designed for multi-class classification tasks, was utilized to measure the discrepancy between the predicted gas classes and the actual labels in the training data. Based on these parameters, the deep model was trained to learn the patterns and relationships between the sensory gas data and their classes. To ensure the preservation of the most accurate model, we employed the Keras callback function, ModelCheckpoint. This function plays a crucial role by saving the model with the highest accuracy achieved and comparing it with the accuracy of each epoch. By monitoring the accuracy, we can determine if the model's performance improves over time. When the accuracy increases, we overwrite the previously saved model with the updated one, thus retaining the best-performing model.
\section{Results and Discussion}
In this section, we present the results and discussion of our study, beginning with a detailed description of the dataset used for training and evaluation. Subsequently, we analyze and discuss the obtained results, focusing on the performance metrics, such as accuracy and loss, and CMs. 

\subsection{Dataset Description}
We used data from Air Quality Dataset for ADL Classification. This dataset consists of a comprehensive collection of indoor gas concentration variations that have been monitored and recorded over time.

The primary purpose of this dataset is to utilize the recorded information to assess and determine the specific types of activities performed within a room or household environment. Thanks to the use of artificial intelligence, a quantitative approach in determining the gas concentration was avoided, which would have required careful calibration of the sensors. 

The dataset contains the values acquired by an array of 6 low-cost sensors in successive instants of time, and the stored values are associated with the particular action that generated them. Through an appropriate data processing, based on one of learning algorithms, after an initial training phase it is possible to recognize the actions that are carried out inside the home. The presence of chemicals in the air is determined through a series of electrochemical gas sensors that have been selected based on the stated technical specifications on the ability to detect classes of compounds. The sensor set can be grouped into two main categories:

\begin{itemize}
  \item MQ Sensors: such as MQ2, MQ9, MQ135, MQ137, MQ138 which have great sensitivity, low latency and low cost; each sensor can respond to different gases.
  \item Analog CO2 gas Sensor: MG-811 which has excellent sensitivity to carbon dioxide and is scarcely affected by the temperature and humidity of the air.
\end{itemize}

The dataset contains 1845 collected samples devided into 4 different types of activities :

\begin{enumerate}
    \item Normal Situation Activity: clean air, a person sleeping or studying or resting.
    \item Preparing Meals Activities: cooking meat or pasta, fried vegetables. One or two people in the room, forced air circulation.
    \item  Presence of Smoke Activity: burning paper and wood for a short period of time in a room with closed windows and doors.
    \item Cleaning Activity: use of spray and liquid detergents with ammonia and / or alcohol. Forced air circulation can be activated or deactivated.
\end{enumerate}

\begin{figure}[t!]
  \centering
   \includegraphics[width=0.45\textwidth]{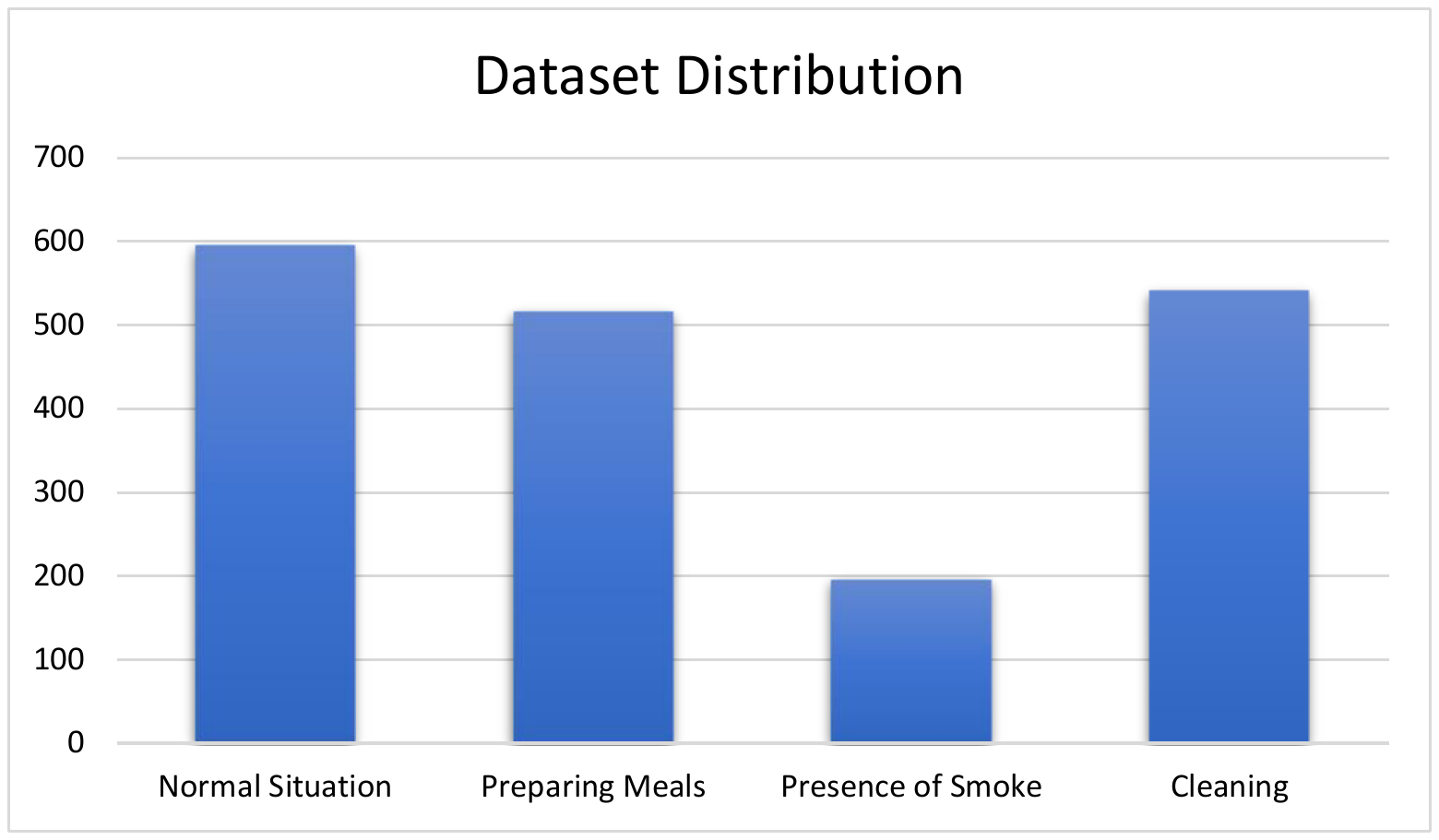}
  \caption{Dataset classes Distribution}
  \label{fig:fig1}
\end{figure}

The dataset distributed as depicted in Fig. \ref{fig:fig1}

Each sample is made up of 7 values: 
\begin{itemize}
    \item The first six values are the sensor outputs.
    \item The last is the label (target or index) of the action generated the values acquired by the sensors.
\end{itemize}

The four different situations are associated with a fairly different composition of the air, taking into account that any activity produces chemical substances due, that is, to human respiration, to the exhalations of metabolic processes, to the release of volatiles by combustion and / or oxidation, and evaporation of household detergents \cite{bengherbia2022real}. After training the proposed model, we obtain the next presented result.

\subsection{Accuracy and Loss Graph}

\begin{figure}[t!]
  \centering
   \includegraphics[width=0.45\textwidth]{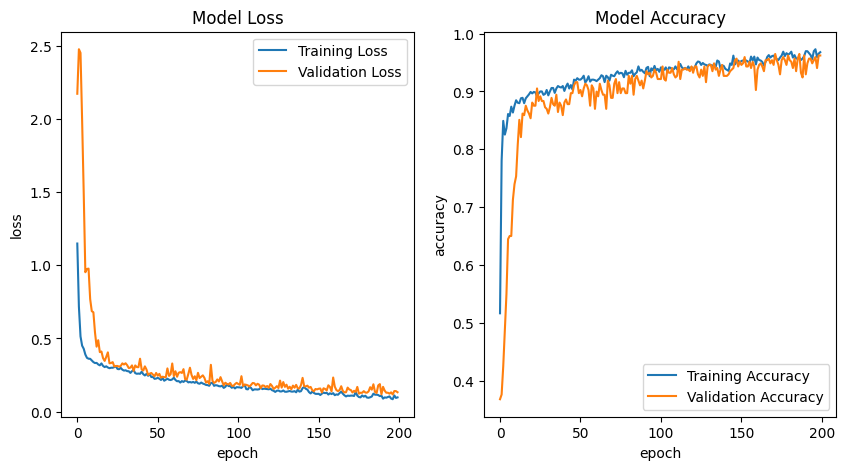}
  \caption{Accuracy and Loss Graph.}
  \label{fig:fig3}
\end{figure}

Fig. \ref{fig:fig3} illustrates a graphical representation of the accuracy and loss metrics for both the training and validation sets as a function of epochs number. 
Notably, both the training and validation accuracy exhibit a proportional increase with the epochs number, indicating an improvement in the model's ability to correctly classify instances. Conversely, the loss metrics show an inversely proportional decrease with the epochs number, signifying a closer alignment of the model's predictions with the actual labels in both training and validation sets. These results suggest that the model effectively learns and adapts its parameters to minimize errors and enhance performance over time.

\subsection{Classification Report}
Table ~\ref{Tab:Tab2} depicts the classification report, providing a comprehensive evaluation of the performance of proposed DL classification model. The classification report includes multiple metrics that offer valuable insights into the overall performance of the model's. These metrics include precision, recall, F1-score and support. Upon analyzing the classification report, it becomes evident that the model’s performance varies across different classes.  Notably, specific observations can be made for the first and second classes :
\begin{itemize}
    \item For the first class (Normal Situation), the accuracy (precision) is reported as 95 percent, indicating relatively lower accuracy rate compared to the other classes. This suggests that the model may encounter challenges in accurately classifying instances belonging to the Normal Situation class.
    \item Regarding the second class (Preparing Meals), the recall is reported as 91 percent, implying that the model may misclassify or overlook instances from this class.
    \item The F1-scores for the first (Normal Situation) and second (Preparing Meals) classes are reported as 97 percent and 95 percent, respectively. The F1-score provides a consolidated assessment of the model's performance, considering both precision and recall.
\end{itemize}
These findings shed light on the strengths and weaknesses of the model across different classes, enabling further analysis and potential enhancements to its classification capabilities of air quality at homes.

\begin{table}[t!]
\centering
\caption{Classification Report}
\begin{tabular}{@{}c|c|c|c|c@{}}
\hline
\textbf{}             & \textbf{Precision} & \textbf{Recall} & \textbf{F1-Score} & \textbf{Support} \\ \hline
\textbf{Normal Situation}            & 95\%               & 100\%           & 97\%              & 122              \\
\textbf{Preparing Meals}            & 99\%               & 91\%            & 95\%              & 108              \\
\textbf{Presence of Smoke}            & 100\%              & 100\%           & 100\%             & 41               \\
\textbf{Cleaning}            & 97\%               & 99\%            & 98\%              & 98               \\ \hline 
\textbf{Accuracy}     &                    &                 & 97\%              & 369              \\
\textbf{Macro Avg}    & 98\%               & 97\%            & 97\%              & 369              \\
\textbf{Weighted Avg} & 97\%               & 97\%            & 97\%              & 369              \\ \hline
\end{tabular}
\label{Tab:Tab2}
\end{table}

\subsection{Confusion Matrix (CM)}
\begin{figure}[t!]
  \centering
   \includegraphics[width=0.30\textwidth]{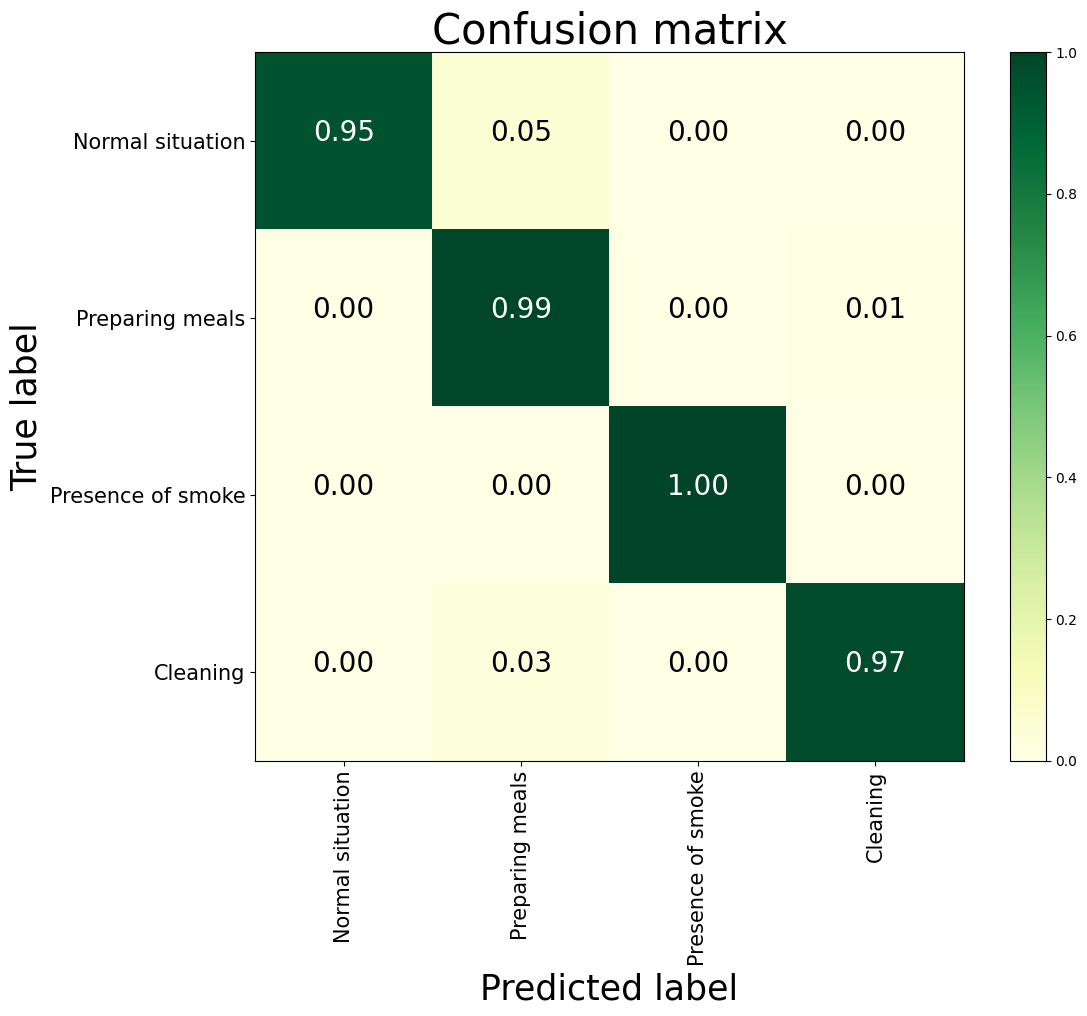}
  \caption{Confusion Matrix of the Training phase}
  \label{fig:fig5}
\end{figure}

The CM is an important tool for evaluating the performance of a classification model, including deep models. It provides a detailed breakdown of the model's predictions compared to the actual labels in a tabular format. 
Based on the analysis of the classification report, it is likely that the similarities between the measurement parameters of the Normal Situation and Preparing Meals classes contribute to the observed classification challenges. This hypothesis can be further supported by examining the CM depicted in Fig. \ref{fig:fig5}.The resemblance between the measurement parameters of the Normal Situation and Preparing Meals classes has an impact on the classification results. Specifically, it is noted that 95 percent of instances belonging to the Normal Situation class are correctly classified as expected. However, there is a misclassification rate of 5 percent, where instances from the Normal Situation class are incorrectly classified as belonging to the Preparing Meals class.These observations highlight the potential confusion that arises due to the similarity between the measurement parameters of these classes, which may require further investigation and potential adjustments in the classification approach to mitigate such misclassifications.We saved the model that achieved the highest accuracy during training for the final testing phase. We evaluate the performance of the final test in terms of accuracy, loss, and prediction time. \ref{Tab:Tab4}.

\begin{table}[t!]
\centering
\caption{The Best Results of CNN 1D Proposed Model}
\begin{tabular}{c|c|c|c}
\hline
\textbf{}           & \textbf{Accuracy (\%)} & \textbf{Loss (\%)} & \textbf{Time (s)}\\ \hline
\textbf{Train}      & 98              & 0.05              & /          \\ 
\textbf{Validation} & 97              & 0.14              & /          \\ 
\textbf{Test}      & 97              & 0.15              & 0.041          \\ 
\hline

\end{tabular}
\label{Tab:Tab4}
\end{table}

\subsection{Final test}

\begin{figure}[t!]
  \centering
   \includegraphics[width=0.30\textwidth]{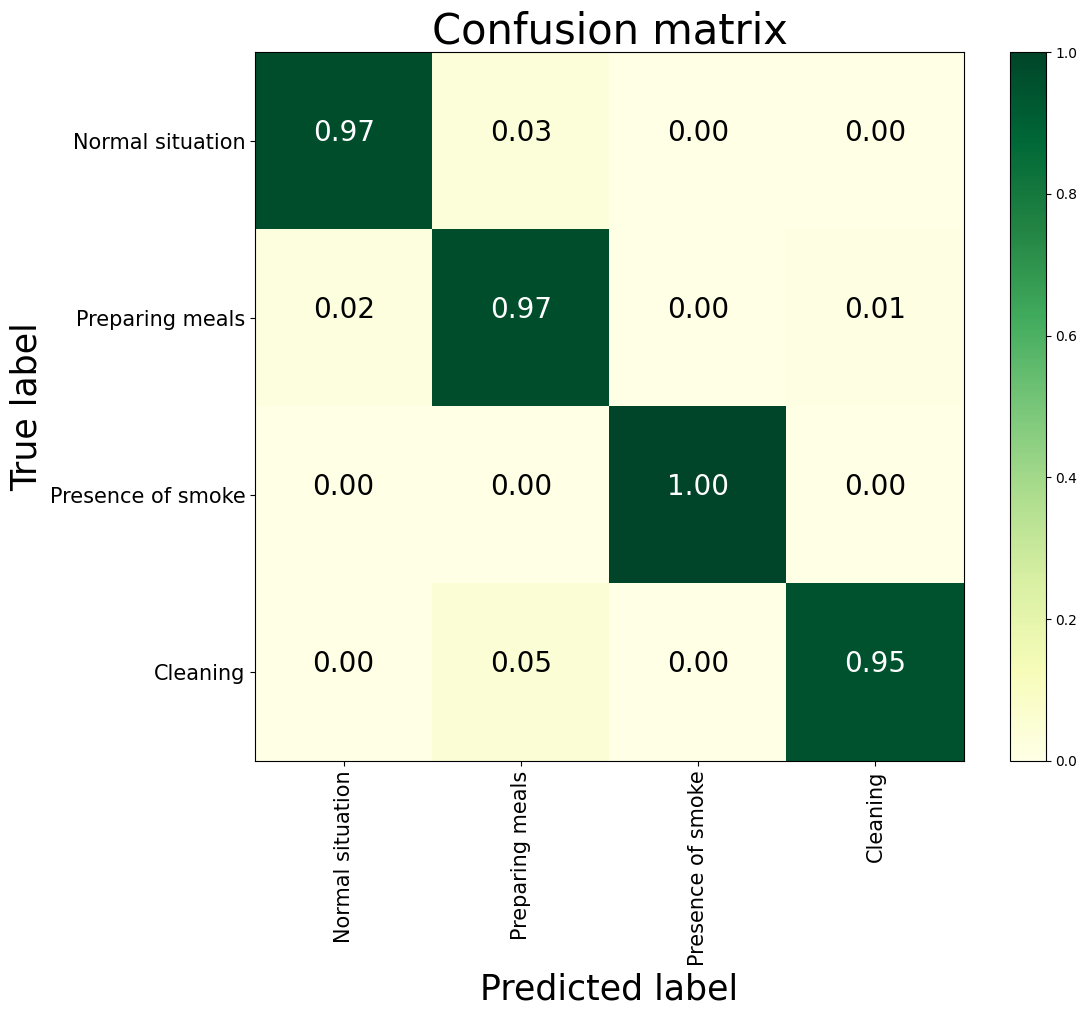}
  \caption{Confusion Matrix of the test phase}
  \label{fig:fig7}
\end{figure}

In this section, we utilize the saved model from the previous phase and assess its performance on the testing set, as illustrated in Table \ref{Tab:Tab4}. The testing set consists of data that were not utilized by the model during the training or validation phases. Consequently, this evaluation provides insights into the model's generalization and performance on unseen data. Fig. \ref{fig:fig7} displays the CM, offering a visual representation of the prediction results for different types of activities based on the final test data, which were not utilized during the model's training or validation stages. The CM summarizes the model's predictions in comparison to the actual labels (target) for each class.
Through analysis of the CM, we observed a recurring issue related to the similarity between the measurement parameters of the first and second classes. This similarity contributes to misclassifications, resulting in confusion between these two classes.




\subsection{Comparison with state of the art}

The best performance of the proposed 1D-CNN model is compared with recent related work using Air Quality Dataset in Table \ref{Tab:Table2}. This comparison demonstrates the effectiveness and reliability of our proposed approach. Accuracy, model size, and prediction time provided are compared. It is worth noting that we were able to significantly improve the accuracy of the 1D-CNN model and reduce the prediction time. The proposed 1D-CNN model achieved A high accuracy of 97.00\% and a prediction time of 41 ms.


\begin{table}[t!]
\centering
\caption{Comparison with the state of the art}
\scriptsize
\begin{tabular}{@{}p{1.1cm}p{0.8cm}p{0.5cm}p{0.4cm}p{0.9cm}p{1cm}p{1.2cm}@{}}
\hline
\textbf{Work}      & \textbf{Model}    & \textbf{Year} & \textbf{Size (ko)} & \textbf{Accuracy (\%)} & \textbf{Prediction Time (s)} & \textbf{Dataset} \\ \hline
E. Gambi et Al ~\cite{gambi2020adl}     & KNN               & 2020          & /                  & 96.32           & /                            & Air Quality          \\
S. Srivatsan et Al ~\cite{srivatsan2022ensemble} & Random Forest     & 2022          & /                  & 96.19           & /                            & Air Quality           \\
Our                & ANN               & 2023          & 120                & 95.90            & 0.043                        & Air Quality           \\
Our                & 1D-CNN            & 2023          & 112                & 97.00            & 0.041                        & Air Quality           \\ \hline
\end{tabular}
\label{Tab:Table2}
\end{table}

\section{Conclusion}



In this paper, we have investigated the application of a CNN 1D DL approach for activity recognition using an IoT air quality sensor dataset. Our objective was to develop a robust system capable of accurately detecting the type of activity based on sensor measurements collected from diverse indoor environments. By leveraging the power of the CNNs in the one-dimensional domain (1D-CNN), we effectively utilized the spatial measurement parameters data. Several experiments were conducted to identify the most accurate model, and the best-achieved results included an accuracy of 97.00 \%, a loss value of 0.15\%, and a prediction time of 41 milliseconds.



\end{document}